\documentclass[12pt,a4paper]{article}

\title{\bf Can a quantum measurement be cancelled in a very short period of time? }
\author{R. L. Schafir\footnote{e-mail:
roger.schafir@londonmet.ac.uk}\\ CISM, London Metropolitan
University, London EC3N 1JY, U.K.}
\date{}
\begin{document}
\maketitle
\begin{abstract}
\noindent A test is suggested for whether the obtaining of certain
information, and then deleting it too quickly to be retained,
constitutes a quantum measurement of that information.
\end{abstract}
\hfil\vskip\baselineskip
\par
The question of what exactly constitutes a measurement, and when
it takes place, remains debatable after more than 75 years of
quantum mechanics. Here a method will be suggested for testing
whether the capture of certain information by macroscopic
instruments, followed by its partial erasure after a very short
period of time, constitutes a measurement of all the information
that was first captured, or only of the information which remains
after the erasure.
\par
Consider a system of two 2-level particles, for instance two
spin-$\frac{1}{2}$ particles, and suppose them in an entangled
state. Then if the particle spins are measured in directions
$\theta $ and $\phi $ respectively, this is a measurement of the
two commuting observables
\begin{equation}
\left (\sigma^A \ _\theta \otimes id , \ id \otimes \sigma^B \
_\phi \right )
\end{equation}
and we can deduce, in the usual way, that measurement of even one
of these observables projects both particles into two definite
unentangled states, and measurements can be made which will
confirm, for instance, the correlations between them.
\par
Suppose instead that we measured
\begin{equation}
\sigma^A \ _\theta \otimes \sigma^B \ _\phi
\end{equation}
This observable represents (as it indicates) the product of the
individual particles' outcomes, and, if the outcomes are treated
as $\pm$1, can be regarded as measuring whether their results are
the same (eigenvalue +1) or opposite (eigenvalue $-$1).
\par
If a measurement is made, choosing either the observables (1) or
the observable (2), the results are compatible, in that the
probabilities of getting the results ``same" or ``opposite" are
the same. But if two successive such measurements are made at
different angles, incompatible results can be
obtained\footnote{This was brought to my attention by G. Adenier
[1], though Adenier argued for a viewpoint which is not adopted
here.}. For example, suppose the two particles are initially in
the (normalized) state
\begin{equation}
\sum_{i,j\:=\:\pm1}\alpha_{ij}\left|i,j\right>
\end{equation}
with respect to z-direction eigenstates, and that, first, both are
measured in the z-direction, and then, second, a measurement is
made of the observables in (1) using the x-direction.  If the
first measurement is of the observables (1), the system is
projected into one of the states $\left|i,j\right>$ with
probabilities $\left|\alpha_{ij}\right|^{2}$, and if both
particles are then measured in the x-direction, each x-eigenstate
is obtained with equal probability, and so in an ensemble of the
two successive measurements there will be equal probability for
each of the four outcomes at the end of the second measurement.
But if the first measurement (in the z-direction) is of the
observable in (2), then the system is projected into one of the
states
\begin{equation}
\frac{1}{\sqrt{p_{1}}}\left\{\alpha_{11}\left|1,1\right>+\alpha_{-1-1}\left|-1,-1\right>\right\},\;\;
\frac{1}{\sqrt{p_{-1}}}\left\{\alpha_{1-1}\left|1,-1\right>+\alpha_{-11}\left|-1,1\right>\right\}
\end{equation}
where $p_{1},p_{2}$ are the respective probabilities, given by
\begin{equation}
p_{1}=\left|\,\alpha_{11}\right|\,^{2}+\left|\,\alpha_{-1-1}\right|\,^{2},\;\;p_{-1}=\left|\,
\alpha_{1-1}\right|\,^{2}+\left|\,\alpha_{-11} \right|\,^{2}
\end{equation}
and then the probability of, for instance, obtaining the
x-eigenstate $\left|1,1\right>_{x}$ at the end of the second
measurement is
\begin{equation}
\frac{1}{4}\left\{\left|\,\alpha_{11}+\alpha_{-1-1}\right|\,^{2}+
\left|\,\alpha_{1-1}+\alpha_{-11}\right|\,^{2}\right\}
\end{equation}
which is clearly $\neq\frac{1}{4}$ (except for special values of
the $\alpha_{ij}$.  It is also generally different to the
probability of $\left|1,1\right>_{x}$ if the first measurement is
omitted, which is
\begin{equation}
\frac{1}{4}\left\{\left|\,\alpha_{11}+\alpha_{-1-1}+
\alpha_{1-1}+\alpha_{-11}\right|\,^{2}\right\}
\end{equation}
\par
If we consider how this could be verified by experiment, the
obvious question is how to measure whether the observers' results
are the same or opposite while leaving the individual results
themselves unknown, and indeed unknowable.\footnote{But see in
this context Ref [2]} But this question can be turned into a test
of quantum measurement. Suppose the particles are brought together
(rather then being measured while spacelike-separated), and the
measurement is made by an apparatus which is considered
macroscopic, and which measures the two outcomes, multiplies them
and retains that result, then irretrievably deletes the original
individual results.  If more than a very short time elapses
between making the individual measurements and deleting them, the
results must be appropriate to a measurement of (1), since the
individual results could have been noted and recorded (whether
they actually were or not).  But if deletion takes place in a time
so short that it was impossible to record the original results,
then we are in the area of ignorance about what exactly
constitutes a quantum measurement, and it is conjectural whether
the results would be appropriate to (1) or to (2).
\par
The situation can obviously be generalised, and does not
necessarily need entangled states.  Suppose $\widehat{A}$ and
$\widehat{B}$ are a complete set of commuting observables on a
system, i.e. $\widehat{A}$ and $\widehat{B}$ are degenerate but
the degeneracy is lifted when both are measured. Suppose the joint
measurement is made, and then, very quickly, a function
$f(\widehat{A},\widehat{B})$ is evaluated from the outcomes, and
recorded by methods which would usually be regarded as classical,
while the values of $\widehat{A}$ and $\widehat{B}$ themselves are
discarded irreversibly.  Provided $f(\widehat{A},\widehat{B})$ is
also degenerate, the projection by $f(\widehat{A},\widehat{B})$ is
not generally the same as projection by the pair
$(\widehat{A},\widehat{B})$.  So would subsequent measurements
give the results expected if the first measurement was of
$(\widehat{A},\widehat{B})$, or $f(\widehat{A},\widehat{B})$?
\par
This article will not attempt to design a specific experimental
apparatus, but the requirements of an apparatus are as follows. We
want the calculation and recording of the value of
$f(\widehat{A},\widehat{B})$ from $(\widehat{A},\widehat{B})$ to
take place by classical means, and we want the subsequent deletion
of $(\widehat{A},\widehat{B})$ to be classical, yet irreversible.
So suppose that the multiplication is carried out by some standard
type of computer or calculator, while the deletion of the original
results takes place before any signal could reach the memory
section of the apparatus, and/or could be retained by the memory
in view of the known volatility and other properties of the
materials used.  Of course it might be that a different design of
apparatus and different materials could change the time threshold
for the deletion, but that would be a different experiment; what
matters is that with the particular design of apparatus which has
been chosen, the information has been discarded irretrievably
before it could be recorded, even if the apparatus was working at
maximum efficiency.
\par
If there is considered to be a difficulty in measuring a
degenerate observable without lifting the degeneracy by
co-measuring a compatible observable (at least accidentally), we
may regard the measurement of $f(\widehat{A},\widehat{B})$ as part
of a non-degenerate measurement one of whose outcomes in each
degeneracy eigenspace, and the only one which can be obtained from
the initial state since the others are orthogonal to it, is the
same state as a direct application of the projector into the
degeneracy eigenspace.  For instance in the entangled particles
example, if the outcomes of projection by $\left (\sigma^A \ _{z}
\otimes \sigma^B \ _{z} \right )$ which are given in (4) are
supplemented by orthogonal states, i.e. by
\begin{equation}
\frac{1}{\sqrt{p_{1}}}\left\{-\alpha_{-1-1}\left|1,1\right>+\alpha_{11}\left|-1,-1\right>\right\},\;\;
\frac{1}{\sqrt{p_{-1}}}\left\{-\alpha_{-11}\left|1,-1\right>+\alpha_{1-1}\left|-1,1\right>\right\}
\end{equation}
we have a basis for a degeneracy-lifting co-measurement of $\left
(\sigma^A \ _{z} \otimes \sigma^B \ _{z} \right )$ which gives the
same outcomes, with the same probabilities, as a measurement of
$\left (\sigma^A \ _{z} \otimes \sigma^B \ _{z} \right )$ on its
own.
\par
If this experiment, or more generally an experiment for
$f(\widehat{A},\widehat{B})$ versus $(\widehat{A},\widehat{B})$,
showed that the second measurement gave results appropriate to the
first measurement being of $f(\widehat{A},\widehat{B})$, despite
the fact that the only way $f(\widehat{A},\widehat{B})$ could have
been evaluated was from the lost results of
$(\widehat{A},\widehat{B})$ (nor could the recorded value of
$f(\widehat{A},\widehat{B})$ be simply a mistake, with no first
measurement having really happened in the circumstances), it must
mean that what was measured was somehow relative to events at the
macroscopic level.  This would be compatible with the Copenhagen
view that a quantum situation resides partly in the macroscopic
context, and is also consistent with the ``irreversible record"
viewpoint that a measurement takes place when a record is created
by an irreversible process.  If the second measurement showed that
the first measurement was of $(\widehat{A},\widehat{B})$, this
would be consistent with ``realist" pictures, decoherence, and
also the many-worlds interpretation, since a splitting of
observers should take place when $(\widehat{A},\widehat{B})$ is
measured, with no splitting for the classical process of
evaluating $f(\widehat{A},\widehat{B})$ and discarding
$(\widehat{A},\widehat{B})$.  Which of the two possibilities would
be found seems to be an open question.

\end{document}